\documentclass[letterpaper, 10 pt, conference]{ieeeconf} % Comment this line out
\IEEEoverridecommandlockouts                             % This command is only
\usepackage{amsmath} % assumes amsmath package installed
\usepackage{amssymb}  % assumes amsmath package installed
\usepackage{subcaption}
\usepackage{textcomp}

\newtheorem{myrem}{Remark}

\usepackage{graphicx}
\usepackage{textcomp}
\usepackage{float}
\usepackage{multirow}
\usepackage{url} 

\PassOptionsToPackage{inline, shortlabels}{enumitem}
\let\labelindent\relax
\usepackage{enumitem}
\newcommand*{\QEDA}{\hfill\ensuremath{\blacksquare}}

\usepackage[linesnumbered,boxed,ruled,commentsnumbered]{algorithm2e}
\SetKwRepeat{Do}{do}{while}%
\usepackage{color}
\usepackage{tikz}
\usetikzlibrary{arrows.meta, positioning}
\usepackage{graphicx}

\title{\LARGE \bf
A Coordinated Routing Approach for Enhancing Bus Timeliness and Travel Efficiency in Mixed-Traffic Environment
}

\author{Tanlu Liang, Ting Bai, and Andreas A. Malikopoulos,~\IEEEmembership{Senior Member, IEEE} % <-this % stops a space
\thanks{This research was supported in part by NSF under Grants CNS-2401007, CMMI-2348381, IIS-2415478, and in part by MathWorks.}\vspace{1.5pt}% <-this % stops a space
\thanks{T. Liang, T. Bai, and A. A. Malikopoulos are with the School of Civil $\&$ Environmental Engineering, Cornell University, Ithaca, New York, USA. A. A. Malikopoulos is also with Systems Engineering Program, Cornell University, Ithaca, New York, USA. E-mails: \{{\tt\small tl933, tingbai, amaliko\}@cornell.edu}}}
\begin{document}

\maketitle
\thispagestyle{empty}
\pagestyle{empty}

%%%%%%%%%%%%%%%%%%%%%%%%%%%%%%%%%%%%%%%%%%%%%%%%%%%%%%%%%%%%%%%%%%%%%%%%%%%%%%%%
\begin{abstract}
This paper proposes a coordinated routing approach that investigates the use of connected and automated vehicles (CAVs) in dedicated bus lanes. The aim is to improve bus schedule adherence while enhancing the travel efficiency of CAVs during the transitional phase of mixed traffic environments. Our approach utilizes real-time traffic data to dynamically reroute CAVs in anticipation of congestion. By continuously monitoring traffic conditions on dedicated lanes and tracking the real-time positions of buses, the system adjusts CAV routes in advance to avoid potential interference with operating buses. This cooperation reduces CAV travel times and minimizes delays that impact transit services. The proposed strategy is validated using microscopic traffic simulations in SUMO. The results demonstrate significant improvements in both transit on-time performance and CAV travel efficiency across a range of traffic conditions.
\end{abstract}

%%%%%%%%%%%%%%%%%%%%%%%%%%%%%%%%%%%%%%%%%%%%%%%%%%%%%%%%%
%-----------Section I. Introduction
\section{Introduction}
Connected and automated vehicles (CAVs) have emerged as a central element in the evolution of modern transportation systems. Advances in vehicle-to-vehicle, vehicle-to-infrastructure, and vehicle-to-everything communication technologies highlight the significant potential of CAVs in improving traffic efficiency, reducing energy consumption, and enhancing road safety~\cite{Malikopoulos2020,9976309,le2024stochastic}. While the adoption of CAVs is projected to grow substantially over the coming decades, the transition towards a fully autonomous fleet is anticipated to unfold gradually~\cite{mallozzi2019autonomous,10209062}. Understanding how to enhance the traffic efficiency under increasing penetration rates of CAVs is therefore critical \cite{Zhao2018CTA,mahbub2023_automatica, Nishanth2023AISmerging}. Extensive research and experimental studies have shown that heterogeneous traffic environments involving CAVs can significantly reduce travel times \cite{zhao2019enhanced,11312531,chalaki2021CSM}. Among various approaches, implementing suitable right-of-way management is pivotal for ensuring the seamless integration of CAVs into existing transportation networks and for fully leveraging their benefits throughout this transition. 

CAV dedicated lane (DL) has been proposed as an effective strategy to enhance the benefits that CAVs can bring to transportation systems \cite{seilabi2023robust}. By providing a separate driving environment, DLs enable CAVs to operate with shorter reaction times and smaller headways, thus increasing lane capacity by up to three times compared to general-purpose lanes (GPLs)~\cite{chen2017towards}. However, in urban settings, the substantial construction costs and limited spatial availability often constrain the feasibility of implementing new CAV DLs~\cite{mehrotra2024construction}. In this context, converting an existing GPL into a CAV DL tends to be a more practical approach. However, it may lead to substantial delays for other types of vehicles using GPLs due to the reduced road capacity~\cite{liu2021strategic}.

To address the tradeoff inherent in implementing DLs for CAVs, a promising strategy has emerged: the CAV and bus \textit{joint} DL. Unlike conventional approaches that require constructing new lanes or displacing human-driven vehicles (HVs), joint use of DLs by buses and CAVs repurposes existing DLs, which were originally designed for buses to ensure their on-time arrivals and maintain public transit efficiency~\cite{xiu2021investigation}. In the existing literature, Chen et al. proposed a space-time modeling framework demonstrating that the mixed use of DLs can enhance CAV travel efficiency and alleviate overall traffic congestion~\cite{chen2020modeling}. More recently, schedule-based joint optimization approaches have been presented to coordinate CAV speeds and bus timetables in Bus Rapid Transit corridors~\cite{Yang2024BRTAutonomous}. In addition, dynamic management strategies have been developed to allocate right-of-way between buses and CAVs through lane-level trajectory planning~\cite{xu2024dynamic}. While these studies provide important insights into joint DL operations, they primarily focus on lane-level coordination or schedule-based optimization, and remain limited in dynamically regulating CAV routing at the network level, particularly in balancing congestion mitigation with reliable bus operations under time-varying demand.

In this paper, we develop an efficient dynamic routing approach to coordinate CAVs traveling in a mixed traffic network, allowing for the DL sharing with buses. The goal is to prioritize the buses' adherence to their scheduled timeliness while taking advantage of the joint DLs for CAVs to improve road capacity, reduce congestion, and enhance system-wide efficiency. By leveraging real-time traffic data, the proposed model enables estimation of potential tense flow on network edges and dynamically navigates the rerouting of CAVs to mitigate congestion on DLs. In summary, the main contributions of this work are twofold. (i) We model the rerouting problem of CAVs in a mixed traffic network where buses and CAVs jointly use DLs to fully take advantage of DLs while enhancing buses' adherence to schedules. (ii) We incorporate a dynamic route condition estimation scheme in the rerouting framework, which enables real-time monitoring of traffic flow and timely adjustment of CAV routes when potential DL congestion is detected. Simulation performed in the Simulation of Urban MObility (SUMO) environment based on realistic traffic network data of San Francisco demonstrates the effectiveness of the proposed approach, indicating feasibility and significant improvements in traffic efficiency through coordinated use of joint DLs among buses and CAVs.

%%%%%%%%%%%%%%%%%%%%%%%%%%%%%%%%%%%%%%%%%%%%%%%%%%%%%%%%%
%---------------Section II. 
\section{Problem Statement}
\begin{figure}[t]
    \centering
    \includegraphics[width=0.4\textwidth]{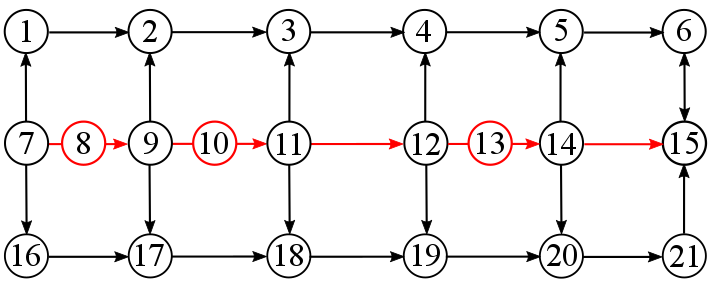}
    \caption{An example illustrating the road network, where bus stations (\(\mathcal{S}\)) are shown by red nodes, intersections (\(\mathcal{I}\)) are shown in black, and DLs are shown by red edges.}
    \label{fig:1}
\end{figure}

Consider an urban road network represented by a directed graph \( G\!=\!(\mathcal{V}, \mathcal{E}) \), where \( \mathcal{V} \) denotes the set of nodes and \( \mathcal{E}\!\subseteq\!\mathcal{V} \!\times\! \mathcal{V} \) denotes the set of directed edges. As illustrated in Fig.~\ref{fig:1}, the node set \( \mathcal{V} \) is comprised of two subsets, denoted as \( \mathcal{V}\!=\!\mathcal{ S } \cup \mathcal{I} \), where \( \mathcal{ S } \) represents the set of bus stations and \( \mathcal{I} \) represents the set of road intersections. Each directed edge \( (v, v') \in \mathcal{E} \) models a road segment from node \( v \) to node \( v' \), with \( v, v' \in \mathcal{V} \). Consider each edge consisting of two lanes in the same direction, i.e., joint DL and GPL, where the GPL can be used by both CAVs and HVs, while DLs are reserved only for joint use by buses and CAVs.

\begin{figure}[t]
    \centering
    \includegraphics[width=0.43\textwidth]{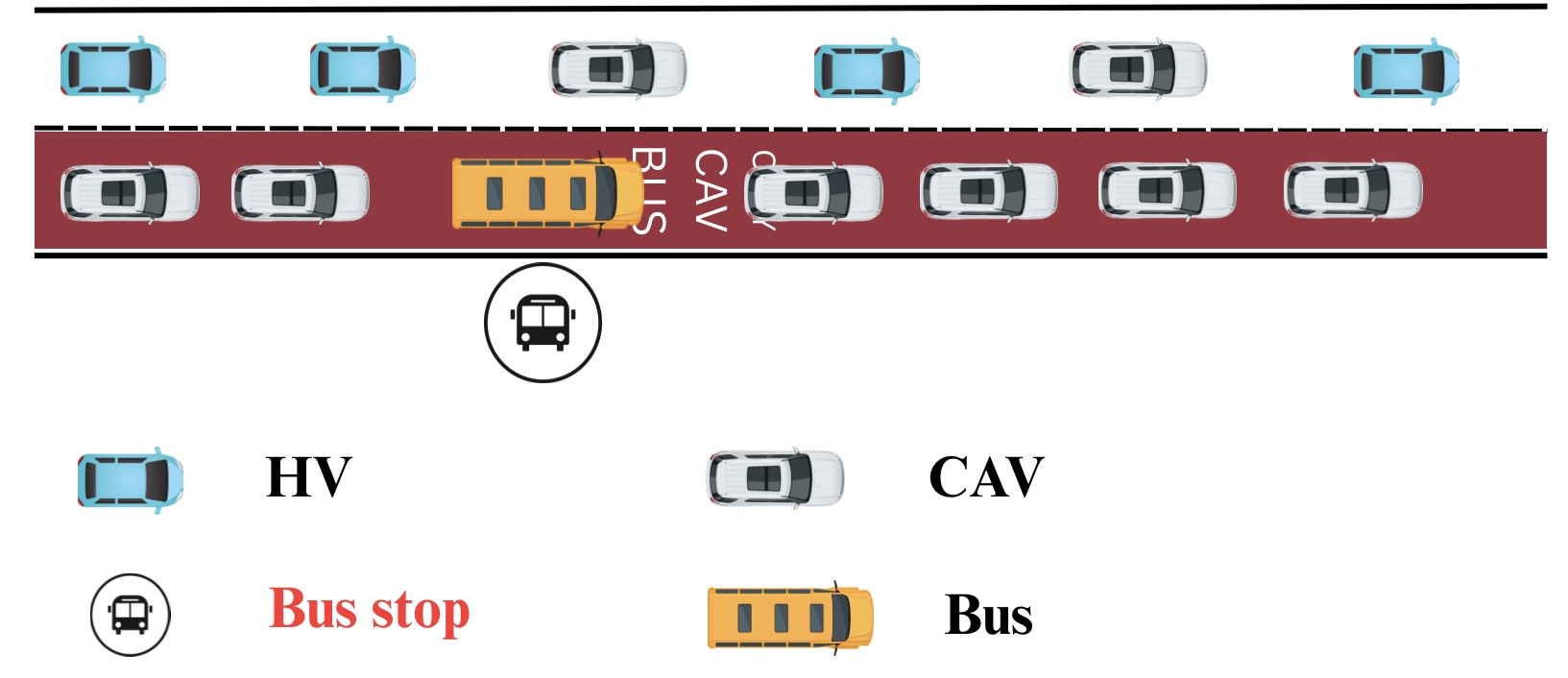}
    \caption{A figure illustrating vehicle types, DLs, and GPLs.}
    \label{fig:2}
\end{figure}

We consider three types of vehicles, including HVs, CAVs, and buses. The sets of HVs, CAVs, and buses within the monitoring time horizon are denoted by $\mathcal{N}^{\text{hv}}$, $\mathcal{N}^{\text{cav}}$, and $\mathcal{N}^{b}$, respectively. Accordingly, the set of all vehicles is represented as $\mathcal{N}\!=\!\mathcal{N}^{\text{hv}}\!\cup\!{\mathcal{N}^{\text{cav}}}\!\cup\!{\mathcal{N}^{b}}$. As given in Fig.~\ref{fig:2}, buses use only DLs and follow a specific timetable. In contrast, HVs can use only GPLs for travel purposes and route planning. With higher flexibility, CAVs can switch between joint DLs and GPLs to improve travel efficiency while minimizing interruptions to buses. 

For any bus $b\in\mathcal{N}^{\text{b}}$, let $P_b$ denote its fixed route, which is represented by a sequence of edges as follows:
\begin{equation}
P_b = \big\{(v_{b,1}, v_{b,2}), (v_{b,2},v_{b,3}), \dots, (v_{b,N_b-1},v_{b,N_b})\big\},  \label{Equ.1}
\end{equation}
where $v_{b,k}\!\in\!\mathcal{I}$ with $k\!=\!1,\dots,N_b$ denotes the $k$th road intersection along the bus route $P_b$, and $N_b$ represents the total number of intersections on the route. We refer to $v_{b,k}^s$ the $k$th bus stop on the edge $(v_{b,k},v_{b,k+1})$.

For each CAV $i\!\in\!\mathcal{N}^{\text{cav}}$ entering the road network, let $(o_i,d_i)$ denote its origin-destination (OD) pair, where $o_i,d_i\!\in\!\mathcal{I}$. Based on real-time traffic conditions, CAV $i$ is assigned an optimal route from its location to the destination. The route of CAV $i$ from $o_i$ to $d_i$ is denoted as
\begin{equation}
    P_{o_i,d_i} = \big\{(o_i,v_k),(v_k,v_{k+1}),\dots,(v_n,d_i)\big\},\label{Equ.2}
\end{equation}
where $v_k,v_{k+1},\dots,v_n\!\in\!\mathcal{I}$ are road intersections in the pre-planned route of the CAV. Note that the route of each CAV is dynamically updated upon its arrival at each intersection in its route. While CAV routes may traverse both intersections and bus stations, re-optimization of the route toward the destination is triggered only when approaching an intersection. HVs are treated as passive participants in the network, following certain routes during their trips.
In this paper, we propose to develop a real-time coordinated routing approach for CAVs that fully leverages dedicated lanes, improves bus schedule adherence, and enhances CAV travel efficiency without creating new obstacles for timely bus arrivals at stops. To this end, a dynamic transportation system is modeled based on real-time traffic flow on each edge. Furthermore, CAVs are selected and rerouted to ensure on-time bus arrivals and maintain traffic efficiency for CAVs. 

\section{Modeling of Traffic Flow}
This section presents the modeling of the road traffic flow in a dynamic transportation network. We start by introducing the traffic flow on joint DLs to capture the potential delays caused by CAVs. Then, the traffic flow on GPLs is modeled for effective rerouting of CAVs.  

\subsection{Traffic Flow on Joint Dedicated Lane}
The travel time for a vehicle to traverse an edge depends on the real-time traffic flow along the edge. Following the classic \textit{Bureau of Public Roads} (BPR) function~\cite{gore2023modified}, let $\tau_e^0$ denote the free-flow travel time to traverse edge $e$. The actual travel time on the edge is then modeled as
\begin{equation}
\tau_e = \tau_e^0 \left( 1 + \alpha \left( \frac{f_e}{c_e} \right)^\beta \right),\label{Equ.3}
\end{equation}
where $f_e$ denotes the real-time flow on $e$, and $c_e$ denotes the effective capacity of the same edge, i.e., the maximum flow that can be sustained without causing congestion. Both $f_e$ and $c_e$ are measured in the number of vehicles per unit time. We assume that all vehicles, regardless of their types, travel at the same speed under free-flow conditions. Additionally, $\alpha$ and $\beta$ are congestion sensitivity parameters. 

Sensors deployed in the road network enable real-time monitoring and analysis of traffic flow patterns~\cite{11312449}. In the road network, we assume that the traffic flow on each edge is measured by sensors located at the starting point of the edge. For each edge on the DLs of bus $b\!\in\!\mathcal{N}^{\text{b}}$, i.e., $(v_{b,k},v_{b,k+1})\!\in\!{P_b}$, we define the traffic flow monitor window 
\begin{equation}
\mathcal{M}(t_{\text{sys}}, \Delta T) = \big[ t_{\text{sys}}\!-\!\Delta T,\; t_{\text{sys}}\!+\!\Delta T \big],
\label{Equ.5}
\end{equation}
where $t_{\text{sys}}$ denotes the system time and $\Delta{T}\in{\mathbb{N}}_{+}$ is a given constant representing the symmetric monitoring window, with $\mathbb{N}_{+}$ denoting the set of positive integers. For each CAV $i\!\in\!\mathcal{N}^{\text{cav}}$ in the road network at time $t_{\text{sys}}$, let $P_i(t_{\text{sys}})$ denote its route planning, which is denoted as
\begin{equation}
P_i(t_{\text{sys}})=\big\{(v_{i,1},v_{i,2}),(v_{i,2},v_{i,3}),\dots,(v_{i,n_i},d_i)\big\},\nonumber
\end{equation}
where $v_{i,\ell}\!\in\!\mathcal{I}$, $\ell\!=\!1,\dots,n_i$, denotes the $\ell$th road intersection along the planned route of CAV $i$ toward its destination. To alleviate traffic congestion on DLs that could lead to delays in bus arrivals at their stops, the cumulative flow of CAVs entering each edge on the DLs is actively monitored. Specifically, let $\tau_{i,\ell}^0$ represent the free-flow travel time on edge $(v_{i,\ell},v_{i,\ell+1})$, and $t_{i,\ell}$ denote the arrival time of CAV~$i$ at intersection $v_{i,\ell}$. The arrival time of CAV $i$ at the next intersection $v_{i,\ell+1}$ is of the form 
\begin{equation}
t_{i,\ell+1}=t_{i,\ell}+\tau_{i,\ell}^0.\label{Equ.6}
\end{equation}
This allows us to compute the accumulated number of CAVs expected to enter the edge $(v_{i,\ell+1},v_{i,\ell+2})$. To facilitate this, we define the following indicated function:
\begin{equation}
I_i(t_{\text{sys}}, \Delta T,v_{b,k})\!=\! 
\begin{cases}
1, & \text{if } t_{i,\ell+1}\!\in\!\mathcal{M}(t_{\text{sys}}, \Delta T),\\
& (v_{i,\ell+1},v_{i,\ell+2})\!=\!(v_{b,k},v_{b,k+1}),\\\\
0, & \text{otherwise}.
\end{cases}\label{Equ.7}
\end{equation}
Then, based on the number of CAVs expected to enter the edge $(v_{b,k},v_{b,k+1})$, the traffic flow on the $k$th edge in the DL of bus $b$ anticipated at time $t_{\text{sys}}$ is computed as
\begin{equation}
\hat{f}_{b,k}(t_{\text{sys}})=\frac{\sum_{i\in\mathcal{N}^{\text{cav}}}I_i(t_{\text{sys}},\Delta T, v_{b,k})}{2\Delta T}.\label{Equ.8}
\end{equation}

\subsection{Traffic Flow on General-Purpose Lane}
Similarly, the time window for traffic flow monitoring at the starting point of each edge on the GPLs is denoted as
\begin{equation}
\mathcal{M}(t_{\text{sys}}, \Delta \tilde{T}) = \big[ t_{\text{sys}}\!-\!\Delta \tilde{T}, t_{\text{sys}}\!+\!\Delta \tilde{T} \big],
\label{Equ.9}
\end{equation}
where $\Delta \tilde{T}\!\in\!\mathbb{N}_{+}$ represents the size of the time window to observe traffic flow. With the knowledge of the route planning of CAVs at time $t_{\text{sys}}$, the anticipated traffic flow on each edge belonging to GPLs is depicted by
\begin{equation}
\!\!\!\hat{f}_{(v,v^{\prime})}(t_{\text{sys}})\!=\!\frac{\sum_{i\in\mathcal{N}^{\text{cav}}}\!I_{(v,v^{\prime})}(t_{\text{sys}},\Delta \tilde{T})\!+\!\Delta N^{\text{hv}}_{(v,v^{\prime})}(t_{\text{sys}})}{2\Delta{\tilde{T}}},\label{Equ.10}
\end{equation}
where the indicated function $I_{(v,v^{\prime})}(t_{\text{sys}},\Delta \tilde{T})$ characterizes the number of CAVs anticipated to traverse the edge $(v,v^{\prime})$ within the monitoring time window, which is given by 
\begin{equation}
I_{(v,v^{\prime})}(t_{\text{sys}}, \Delta \tilde{T})= 
\begin{cases}
1, & \text{if } t_{i,\ell+1}\!\in\!\mathcal{M}(t_{\text{sys}}, \Delta \tilde{T}),\\
& (v_{i,\ell+1},v_{i,\ell+2})\!=\!(v, v^{\prime}),\\\\
0, & \text{otherwise},
\end{cases}\label{Equ.11}
\end{equation}
where $(v,v^{\prime})\!\in\!\mathcal{E}\!\setminus\! \cup_{b\in\mathcal{N}^{\text{cav}}} P_b$ denotes an edge located on the GPLs. Note that $\Delta{N}^{\text{hv}}_{(v,v^{\prime})}(t_{\text{sys}})$ represents the increase in the number of HVs, whose route choices are not controlled by the system planner but can be monitored.  

Subsequently, by the BPR model in \eqref{Equ.3}, the travel time for CAVs to traverse each edge on GPLs can be denoted by
\begin{equation}
\hat{\tau}_{(v,v^{\prime})}(t_{\text{sys}}) = \tau_{(v,v^{\prime})}^0\left( 1 + \alpha \left( \frac{\hat{f}_{(v,v^{\prime})}(t_{\text{sys}})}{c_{(v,v^{\prime})}} \right)^\beta \right),
\label{Equ.12}
\end{equation}
where $\tau_{(v,v^{\prime})}^0$ denotes the free-flow travel time on $(v,v^{\prime})$, and $\hat{f}_{(v,v^{\prime})}(t_{\text{sys}})$ is the anticipated traffic flow computed by \eqref{Equ.10}.

\begin{myrem}
The traffic flow monitoring windows are defined by $\Delta {T}$ in \eqref{Equ.5} for joint DLs and by $\Delta {\tilde{T}}$ in \eqref{Equ.9} for GPLs, respectively, to capture real-time traffic conditions. A larger monitoring window puts greater emphasis on the macroscopic characteristics of traffic flow on network edges. %With a different emphasis, the sizes of the two monitoring windows are not necessarily identical. 
\end{myrem}

\section{Dynamic Rerouting Approach}
In this section, we propose a coordinated rerouting approach that dynamically optimizes route assignments for CAVs based on real-time traffic data. The objective is to ensure that buses adhere to their scheduled arrival times at each stop while enhancing the overall traffic efficiency of CAVs. The proposed optimization scheme consists of identifying the set of CAVs to be rerouted and determining the optimal alternative routes for those CAVs.   

\subsection{Identification of CAVs for Rerouting}
Let $\hat{t}_{b,k}$ denote the estimated arrival time of bus $b\!\in\!\mathcal{N}^{\text{b}}$ at its $k$th road intersection $v_{b,k}$. Upon arriving at its $(k\!-\!1)$th road intersection at time $t_{b,k-1}\!=\!t_{\text{sys}}$, the route ahead is denoted as
\begin{equation}
P_b(t_{\text{sys}})=\big\{(v_{b,k},v_{b,k+1}),\dots,(v_{b,N_b-1},v_{b,N_b})\big\},\label{Equ.13}
\end{equation}
excluding the intersection $v_{b,k-1}$. As such, the estimated arrival time $\hat{t}_{b,k}$ is represented as
\begin{equation}
\hat{t}_{b,k}=t_{b,k-1}+\tau_{b,k-1}^0,\label{Equ.14}
\end{equation}
where $\tau_{b,k-1}^0$ is the free-flow travel time from $v_{b,k-1}$ to $v_{b,k}$.

Recall that the traffic flow $\hat{f}_{b,k}(t_{\text{sys}})$ on edge $(v_{b,k},v_{b,k+1})$ anticipated at time $t_{\text{sys}}$; cf.~\eqref{Equ.8}. Employing the BPR function in \eqref{Equ.3}, the anticipated travel time on the same edge is as the following form:
\begin{equation}
\hat{\tau}_{b,k}(t_{\text{sys}}) = \tau_{b,k}^0\left(1+\alpha\left( \frac{\hat{f}_{b,k}(t_{\text{sys}})}{c_{b,k}} \right)^{\beta} \right).
\label{Equ.15}
\end{equation}
\begin{figure}[t]
    \centering
    \includegraphics[width=0.4\textwidth]{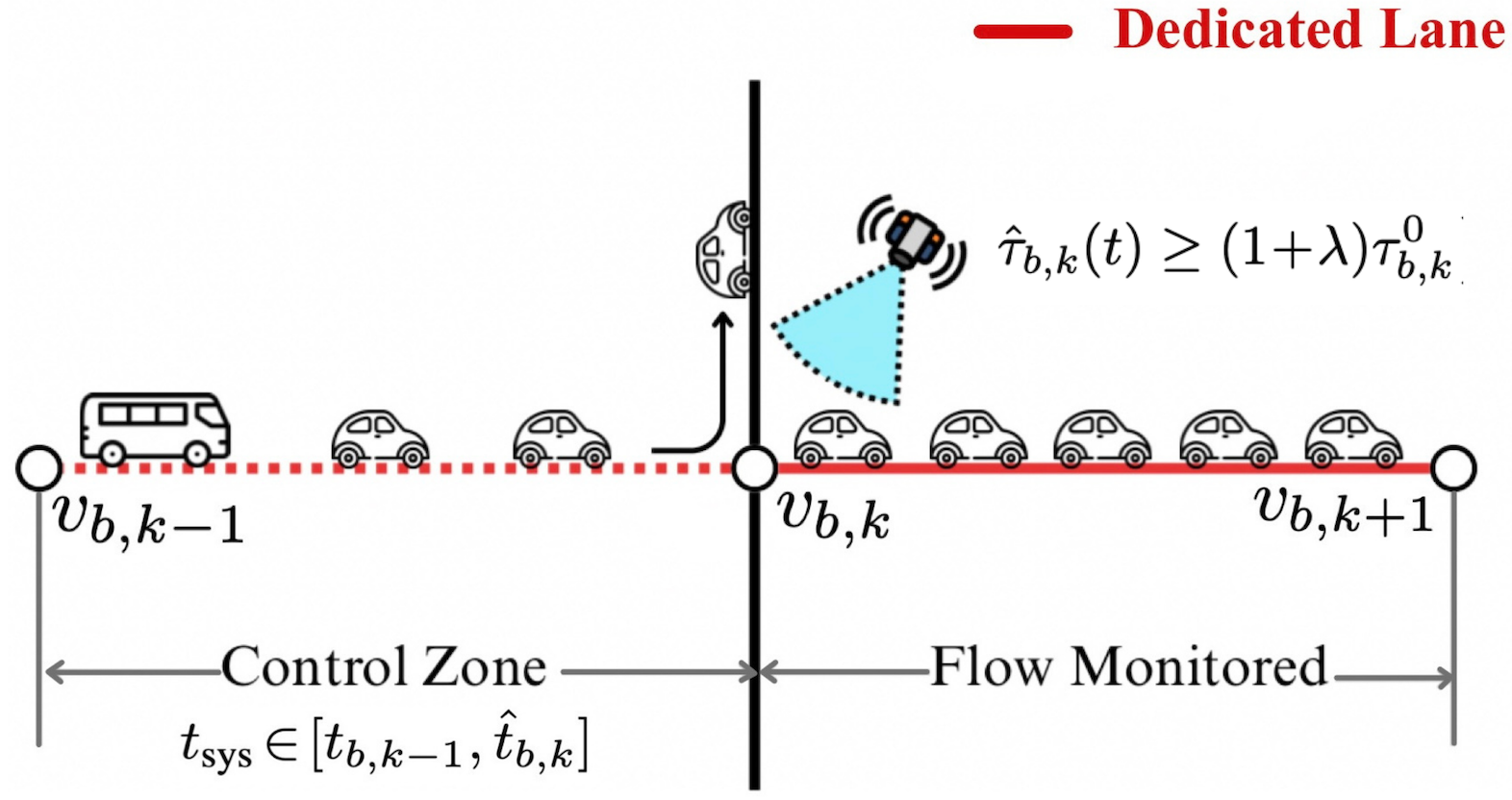}
    \vspace{-2pt}
    \caption{Rerouting scheme at each intersection in DLs.}
    \label{fig:3}
\end{figure} As illustrated in Fig.~\ref{fig:3}, we define the control horizon for bus $b$ as the time interval $[t_{b,k-1},\hat{t}_{b,k}]$. For any system time $t_{\text{sys}}\!\in\![t_{b,k-1},\hat{t}_{b,k}]$, the set of CAVs on edge $(v_{b,k-1},v_{b,k})$ that need to be rerouted, denoted as $\mathcal{R}_{b,k-1}(t_{\text{sys}}^{\prime})$, is expressed as
\begin{equation}
\mathcal{R}_{b,k-1}(t_{\text{sys}}^{\prime})=\big\{i\!\in\!{\mathcal{N}^{\text{cav}}}:I_i(t_{\text{sys}}^{\prime},\Delta{T},v_{b,k})\!=\!1\big\}.\label{Equ.16}
\end{equation}
As defined \eqref{Equ.7}, $I_i(t_{\text{sys}}^{\prime},\Delta{T},v_{b,k})\!=\!1$ indicates that CAV~$i$ is within the monitor window of $v_{b,k}$ at time $t_{\text{sys}}^{\prime}$, which represents the time when congestion on edge $(v_{b,k},v_{b,k+1})$ is detected and CAV rerouting is triggered. Namely, 
\begin{equation}
t_{\text{sys}}' \in 
\left\{ t \in [t_{b,k-1}, \hat{t}_{b,k}] :
\hat{\tau}_{b,k}(t) \ge (1\!+\!\lambda)\tau_{b,k}^0
\right\}
\label{Equ.17}
\end{equation}
where $\lambda\!>\!0$ is a small tolerance.

%\begin{myrem}
%In the developed rerouting approach, the traffic flow on edge $(v_{b,k},v_{b,k+1})$ is dynamically %monitored, and the CAVs on the preceding edge, i.e., $(v_{b,k-1},v_{b,k})$, are controlled before %bus $b$ entering $(v_{b,k},v_{b,k+1})$. This procedure enables effective congestion mitigation on each edge along the DLs. Thus, it is reasonable to employ the free-flow travel time in %\eqref{Equ.14} to estimate the bus's arrival time at $v_{b,k}$.
%\end{myrem}

\subsection{Route Optimization for Rerouted CAVs}
To mitigate traffic congestion on edge $(v_{b,k},v_{b,k+1})$ while improving the traffic efficiency of the transportation network, CAVs identified for rerouting are each assigned a new optimal route when $\hat{\tau}_{b,k}(t_{\text{sys}})\!\geq\!(1\!+\!\lambda){\tau_{b,k}^0}$ is detected. Specifically, for each CAV $i\!\in\!\mathcal{R}_{b,k-1}(t_{\text{sys}}^{\prime})$, the system planner re-optimizes its route from intersection $v_{b,k}$ to $d_i$, using the real-time traffic conditions observed at time $t_{\text{sys}}^{\prime}$. Based on the latest traffic data, including the estimated travel times for each edge on GPLs and those on DLs, excluding the edge $(v_{b,k},v_{b,k+1})$, the optimization problem addressed for optimal route planning is cast as
\begin{align}
&\min \  \sum_{i \in \mathcal{R}_{b,k-1}(t_{\text{sys}}^{\prime})} \sum_{(v_{i,\ell},v_{i,\ell+1})\in{P_i(t_{\text{sys}}^{\prime}})}^{}\!\!\!\!\!\!\! \hat{\tau}_{i,\ell}(t_{\text{sys}}^{\prime})\label{Equ. 18}\\
 & \ \mathrm{s.\,t.} \quad ~P_i(t_{\text{sys}}^{\prime})\!\in\!\mathcal{P}_{(v_{b,k},d_i)}, ~i\!\in\!\mathcal{R}_{b,k-1}(t_{\text{sys}^{\prime}}),\label{Equ.19}\\
 & \quad \quad \quad  ~\eqref{Equ.5}-\eqref{Equ.12},\ \eqref{Equ.14}-\eqref{Equ.17},\nonumber
\end{align}
where $\hat{\tau}_{i,\ell}(t_{\text{sys}}^{\prime})$ represents the anticipated travel time to traverse edge $(v_{i,\ell},v_{i,\ell+1})$, estimated based on traffic conditions at time $t_{\text{sys}}^{\prime}$. As defined previously, $P_i(t_{\text{sys}}^{\prime})$ denotes the route planning of CAV $i$ at time $t_{\text{sys}}^{\prime}$ to complete the remainder of its trip. For CAVs to be rerouted, $P_i(t_{\text{sys}}^{\prime})$ refers to a route from intersection $v_{b,k}$ to the destination $d_i$. We denote by $\mathcal{P}_{(v_{b,k},d_i)}$ the set of all feasible routes connecting $v_{b,k}$ to $d_i$.  

Constraints \eqref{Equ.5}-\eqref{Equ.8}, \eqref{Equ.14}, and \eqref{Equ.15} provide estimates of traffic flow and travel times for each edge on joint DLs, while constraints \eqref{Equ.9}-\eqref{Equ.12} offer travel time estimates for edges on GPLs. Moreover, \eqref{Equ.16} and \eqref{Equ.17} identify the set of CAVs selected for route replanning and specify the triggering condition for rerouting. After all estimated travel times across the network are updated at time $t_{\text{sys}}^{\prime}$, a prediction-aware Dijkstra algorithm~\cite{7978471} is executed to determine the optimal rerouting paths for all identified CAVs, denoted as $P_i^*(t_{\text{sys}}^{\prime})$, where $i\!\in\!\mathcal{R}_{b,k-1}(t_{\text{sys}}^{\prime})$. In this modified Dijkstra framework, the edge weights correspond to the latest updated estimated travel times instead of static free-flow travel times.   
%\begin{myrem}
%Solving the above route optimization problem requires real-time traffic data collected from existing sensor infrastructures, such as loop detectors and cameras. These data are aggregated within monitoring windows to compute vehicles' actual travel times on the current edges and to estimate their travel times on future edges. Incorporating travel time estimations enables long-horizon congestion prediction and route adjustment in advance, thus preventing vehicles from being trapped in severe congestion.   
%\end{myrem}
\begin{myrem}
The initial route of each CAV is computed based on the prevailing traffic conditions at the time the vehicle starts its trip, to minimize its own travel time. When rerouting is triggered for a group of CAVs, the system planner reoptimizes their routes to minimize their total travel time, thus enhancing overall traffic efficiency. This approach accounts for the diverse departure times of CAVs and ensures that the minimal set of CAVs is rerouted only when necessary, balancing CAVs' flexibility, transit reliability, and system-wide efficiency. 
\end{myrem}
%=================================================================
\section{Simulation Study}
This section presents simulation studies conducted on the SUMO platform to evaluate the effectiveness of the proposed approach. The code implementation is available online.\footnote{See code at: {https://github.com/Tanlu-L/Coordinated-Routing-Approach-for-Enhancing-Bus-Timeliness-and-Travel-Efficiency.git}}

\subsection{Simulation Setup}

\begin{figure}[t]
    \centering
    \includegraphics[width=0.45\textwidth]{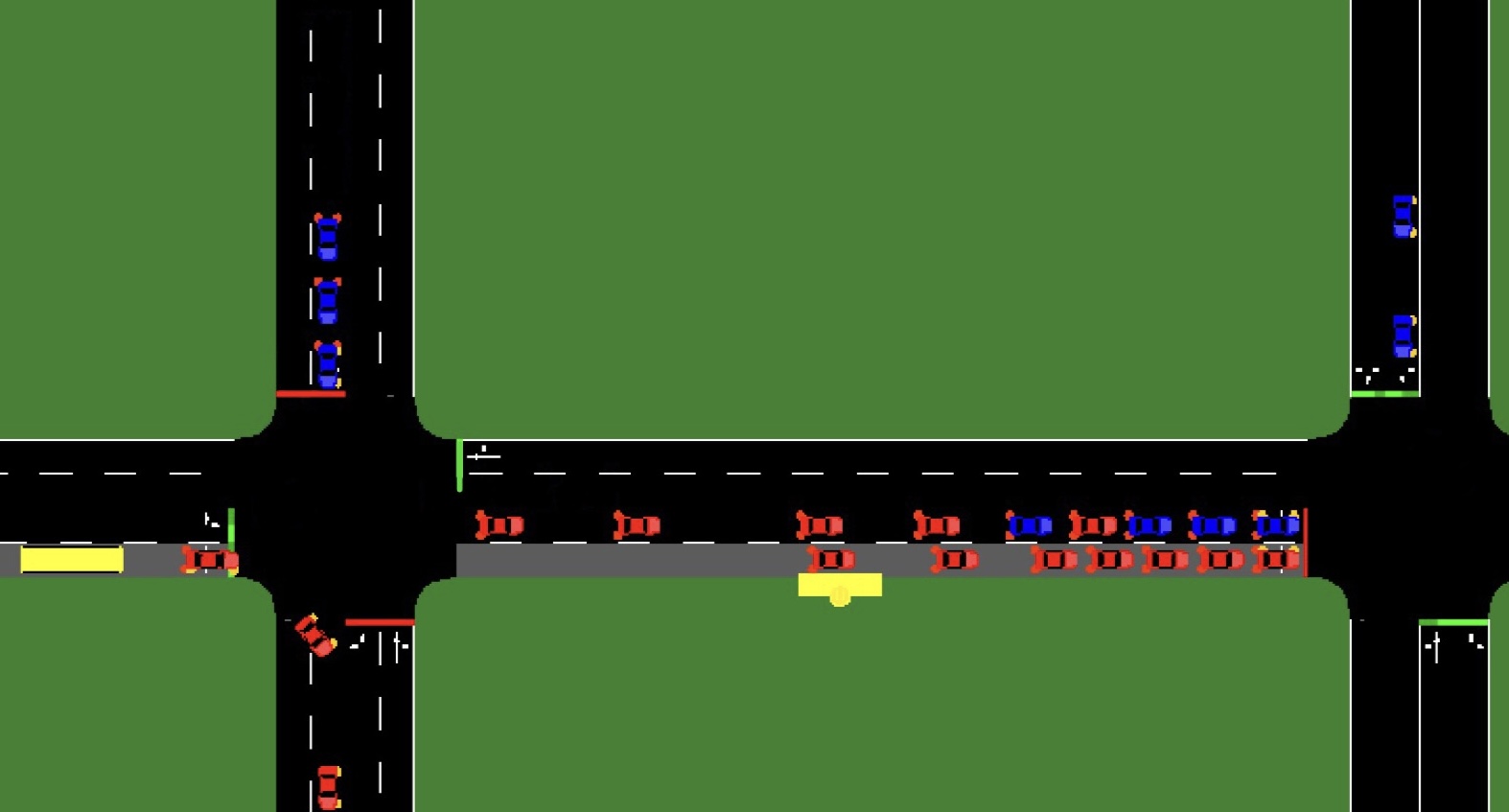}
    \caption{Illustration of the SUMO simulation environment, where the bus is depicted in yellow, CAVs are shown in red, and HVs are shown in blue. DLs are highlighted in grey, with bus stops marked in yellow.}
    \label{fig:4}
\end{figure}
As shown in Fig.~\ref{fig:4}, we construct an urban road network in SUMO with the road topology matching that shown in Fig.~\ref{fig:1}. This network replicates Van Ness Avenue in San Francisco from Market Street to Lombard Street. The roadway geometry and signal settings were obtained from \textit{OpenStreetMap}. This corridor is representative of typical transit-priority urban arterials with DL operations. We consider that buses follow a predefined route along the joint DLs, traveling from the origin (node $7$ in Fig.~\ref{fig:1}) to the destination (node $15$). Sensors are assumed at road intersections, with monitoring windows set to $\Delta T\!=\!30$~s for joint DLs and $\Delta\tilde{T}\!=\!60$~s for GPLs.

\begin{table}[t]
\caption{Scheduled bus arrival times.} % title of the table
\centering
\renewcommand{\arraystretch}{1.4} % increase vertical spacing (1.3–1.5 looks good)
\begin{tabular}{c c c c c}
\hline
Stop & Station 1 & Station 2 & Station 3 & Destination \\ 
\hline
Time [mm:ss] & 01:03 & 03:46 & 07:39 & 09:27 \\
\hline
\end{tabular}
\label{tab:bus_schedule}
\end{table}

We model a peak-hour service with $10$ buses departing every $6$~min from the same origin. The timetable in Table~\ref{tab:bus_schedule} is aligned with the real schedule and generated from free-flow travel times with $60$~s dwell per stop and a $30$~s signal margin. CAVs and HVs enter at rates of $8$ and $20$ veh/min, respectively ($30\%$ CAV penetration). The OD pair for each CAV is set as $(7,15)$, while the OD pair for each HV is randomly selected from the set $\{(1,6), (7,15), (16,21)\}$.

We compare the proposed coordinated routing strategy against two baseline methods: (i) static route planning (SRP) without rerouting and (ii) dynamic route planning (DRP) with rerouting. In the SRP strategy, routes are assigned based on free-flow travel times and remain fixed throughout the trip. In DRP, each CAV is initially assigned the shortest path using real-time traffic data and is rerouted if a lower-travel-time path is detected within a $60$~s monitoring window. In the proposed method, CAVs follow the predictive routing strategy described in Section~IV. For both SRP and DRP, shortest paths are computed using a weighted Dijkstra algorithm, where edge weights are initialized by free-flow travel times and updated dynamically based on real-time traffic conditions.

\subsection{Results and Analysis}

\begin{figure}[t]
    \centering
    \includegraphics[width=0.46\textwidth]{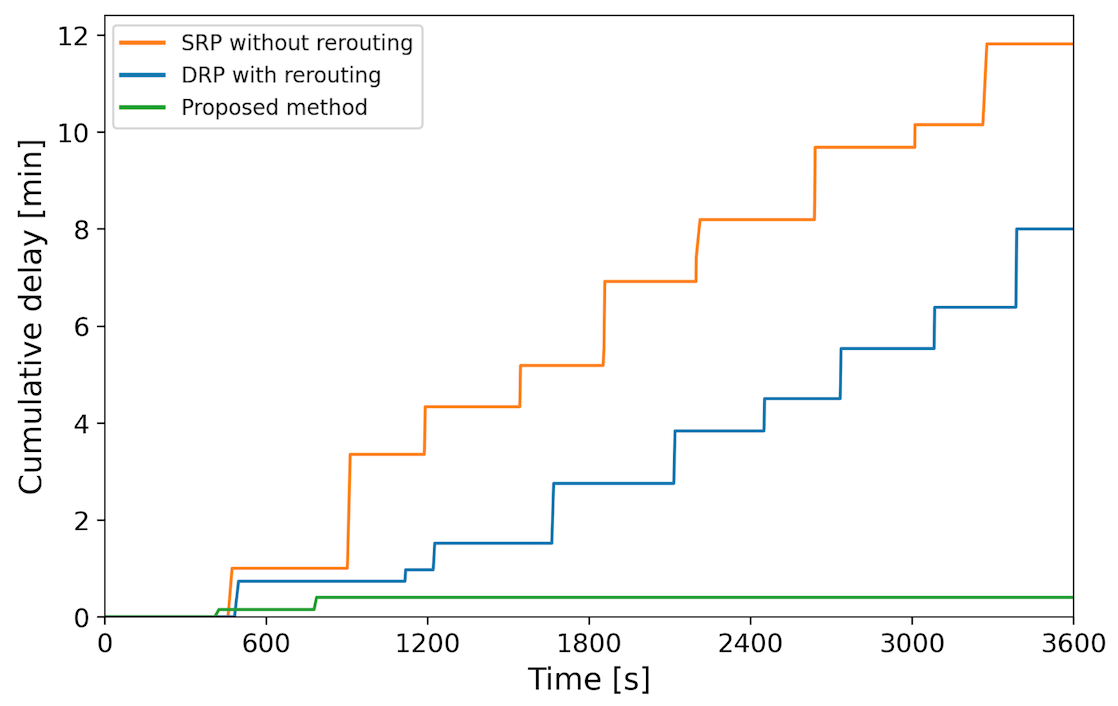}
    \vspace{-3pt}
    \caption{The dynamics of the accumulated bus delay.}
    \label{fig:5}
\end{figure}

\begin{table}[t]
\centering
\caption{Comparison of bus on-time percentage.}
\label{tab:station_perf}
\setlength{\tabcolsep}{10pt}       % adjust column spacing
\renewcommand{\arraystretch}{1.3}  % adjust row spacing
\footnotesize
\begin{tabular}{cccc}
\hline
 Methods & Station 1 & Station 2 & Station 3 \\
\hline
\quad SRP using joint DL   & $10\%$ & $30\%$ & $30\%$ \\
\quad DRP using joint DL   & $50\%$ & $60\%$ & $60\%$ \\
\quad Proposed method      & $90\%$ & $90\%$ & $90\%$ \\
\hline
\end{tabular}
\end{table}

We evaluate the accumulated bus delay, defined as the total additional travel time experienced by all buses relative to their scheduled arrival times upon arrival at the destination, over a 3,600-second simulation horizon. The comparative results are presented in Fig.~\ref{fig:5}. As shown, buses experience significant delays under SRP, while DRP provides only partial improvement. In contrast, the proposed method nearly eliminates bus delays by integrating real-time traffic conditions and potential congestion estimation with efficient CAV rerouting. Table~\ref{tab:station_perf} reports the on-time percentage of the $10$ buses at each station, defined as arrivals within $60$~s of the scheduled times in Table~\ref{tab:bus_schedule}. The SRP, DRP, and proposed method achieve on average $23\%$, $57\%$, and $90\%$ on-time performance, respectively, demonstrating the advantage of the proposed approach.

\begin{figure}[t!] \centering \includegraphics[width=0.46\textwidth]{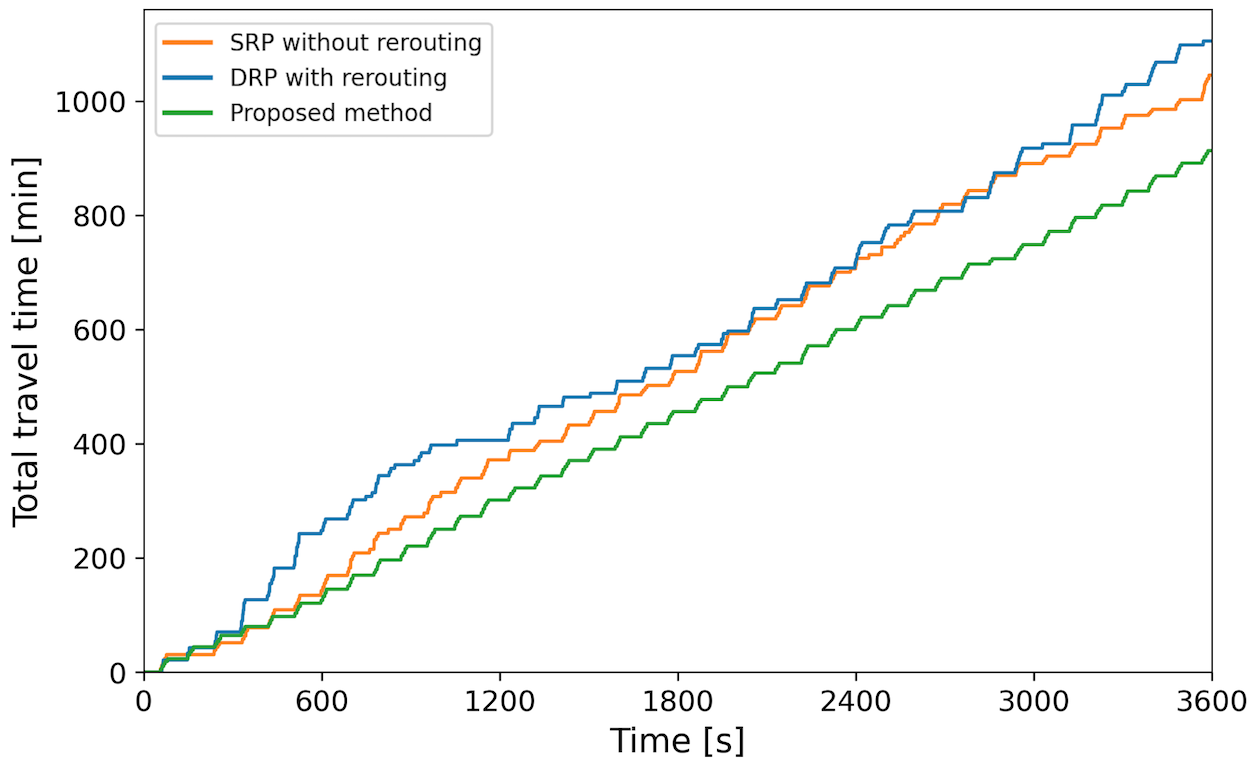} \caption{The dynamics of the cumulative CAV travel time.} \label{fig:6} \end{figure}

\begin{figure}[t!] \centering \includegraphics[width=0.46\textwidth]{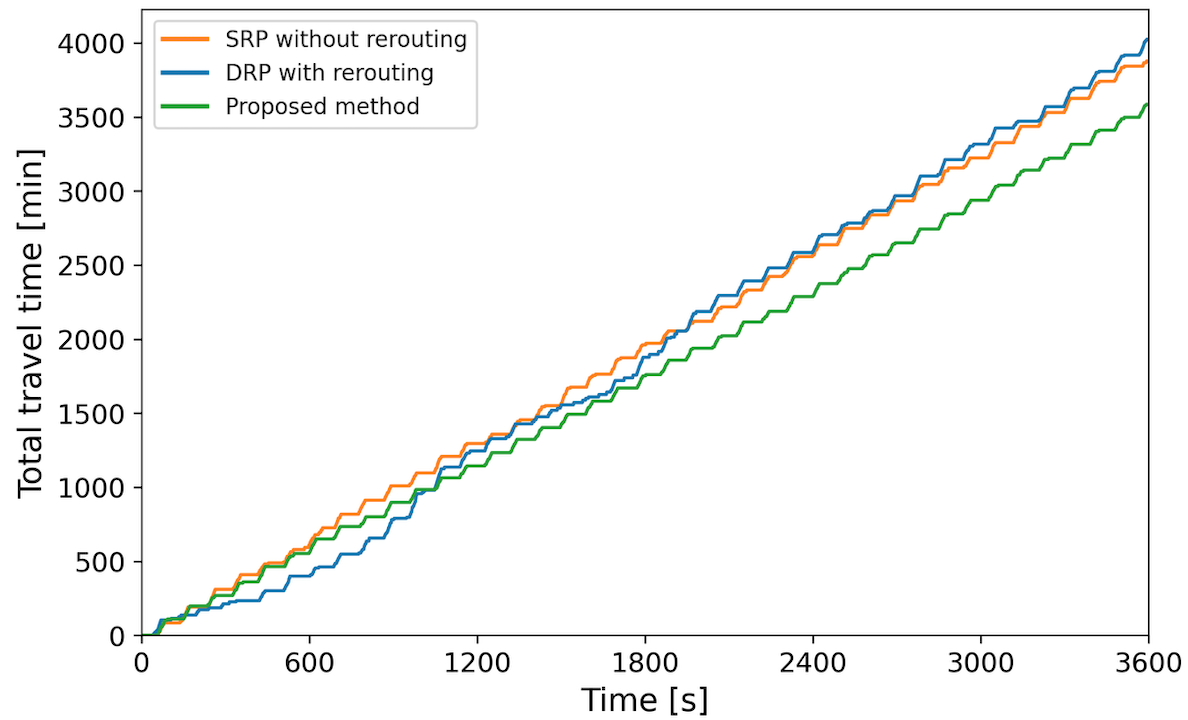}  
 \vspace{-2pt}\caption{The dynamics of the cumulative HV travel time. } \label{fig:7} \end{figure}

Fig.~\ref{fig:6} and Fig.~\ref{fig:7} show the total travel time of CAVs and HVs over the simulation horizon. The proposed method achieves the lowest overall travel time for both vehicle classes by proactively diverting CAVs from congestion-prone edges, particularly those affected by bus dwell events, thereby improving network flow continuity. Note that the DRP approach yields longer CAV travel times than both SRP and the proposed method, a common effect in dynamic routing where vehicles simultaneously shift to shorter paths and induce secondary congestion.%Fig.~\ref{fig:9} shows the accumulated delay for each bus across all stations, indicating consistent delay reduction under the proposed strategy. Note that bus~$6$ experiences relatively higher delay due to heavier traffic conditions, while buses entering earlier or later encounter less congestion. Such a pattern also aligns with the observed distribution of CAVs and HVs volumes in the system.

%\begin{figure}[t]
%    \centering
%    \includegraphics[width=0.42\textwidth]{Fig.9.png}
%    \vspace{-5pt}
%    \caption{The accumulated bus delay at experienced stops.}
%    \label{fig:9}
%\end{figure}      
\begin{figure}[t]
    \centering
    \includegraphics[width=0.41\textwidth]{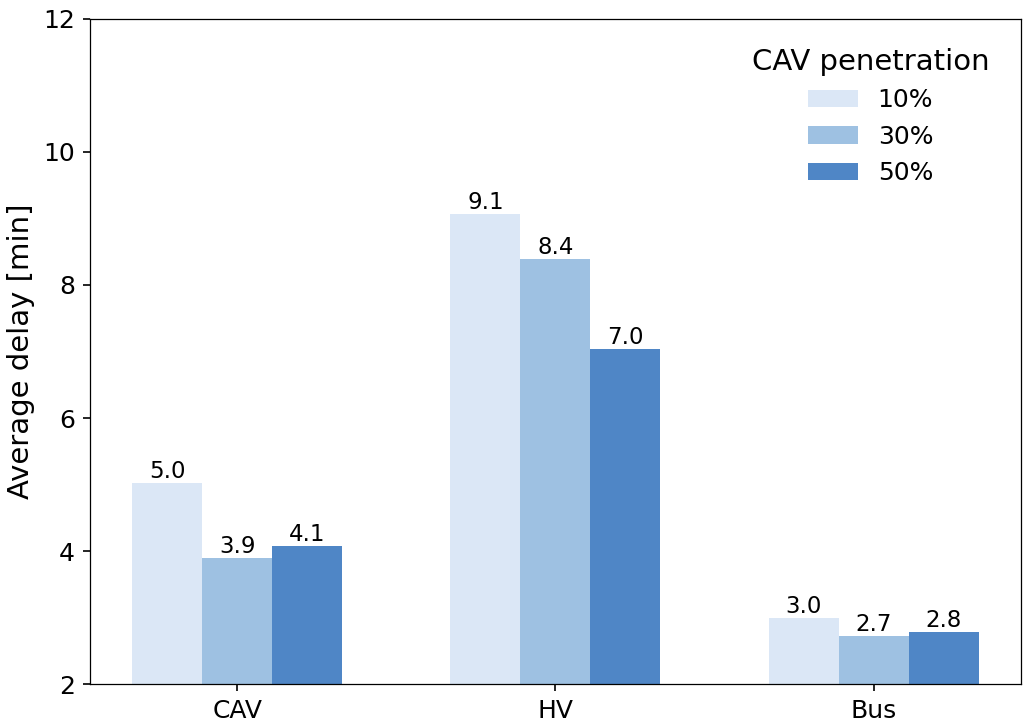}
    \caption{The trip delay in different CAV penetration.}
    \label{fig:10}
\end{figure}
We further evaluate the efficiency of the proposed method under different CAV penetration rates. Fig.~\ref{fig:10} presents the average trip delays experienced by buses, HVs, and CAVs under penetration levels of $10\%$, $30\%$, and $50\%$, with the total travel demand set to $2,000$ vehicles per hour. The results show that HV delays steadily decrease as CAV penetration increases, while CAV and bus delays exhibit little variation between $30\%$ and $50\%$ penetration. These findings show that the benefits of increased penetration saturate beyond a certain threshold, with limited additional travel time savings. 

\begin{figure}[t]
    \centering
    \includegraphics[width=0.42\textwidth]{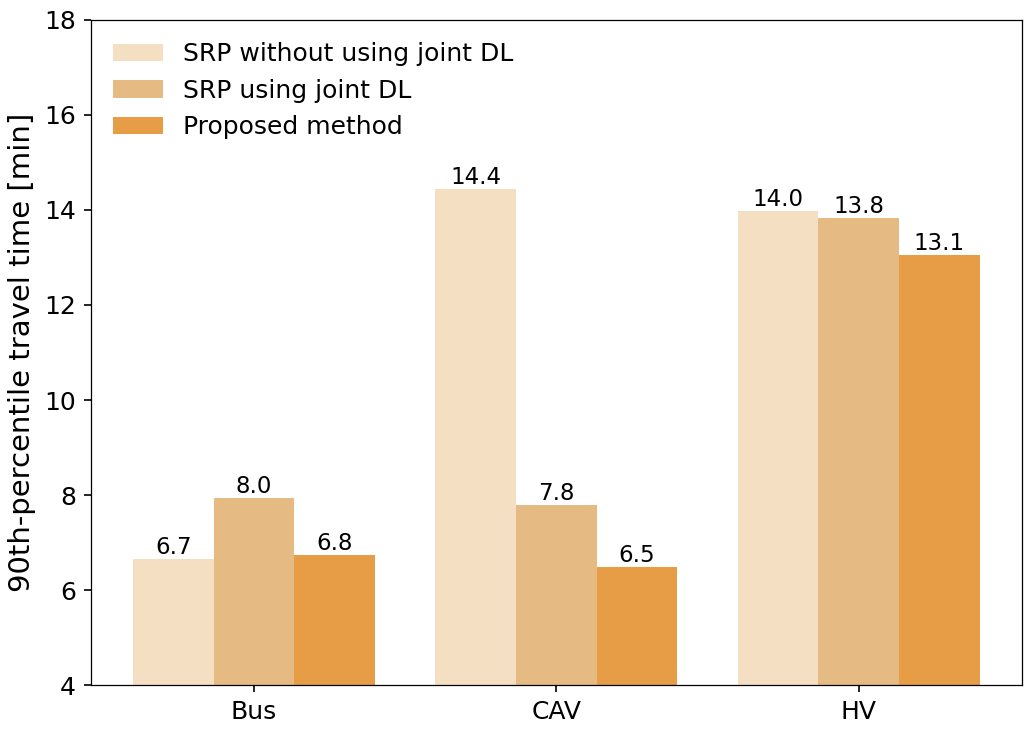}
    \caption{90th-percentile travel time comparison.}
    \label{fig:11}
\end{figure}
To further evaluate the efficiency of the proposed method, we introduce a scenario termed SRP without joint DL access. In this setting, we emulate a dedicated bus lane by allowing only buses to use the DL, while CAVs follow their initially assigned shortest routes. Fig.~\ref{fig:11} shows the 90th-percentile travel time for each vehicle type under SRP without joint DL, SRP with joint DL, and the proposed method. The comparison reveals that allowing CAVs to share the joint DL reduces travel times for both CAVs and HVs, but at the expense of increased bus travel time. In contrast, the proposed method maintains bus travel time close to that of the dedicated-lane scenario while further reducing CAV and HV travel times compared with SRP using joint DL.

Given the above results, we validate the effectiveness of the proposed rerouting strategy in preserving bus transit priority and improving schedule adherence. Through the joint use of DLs with buses, both CAVs and HVs benefit from the dynamic rerouting with reduced trip delays. These findings demonstrate that the congestion-aware coordinated routing approach improves travel efficiency across the entire system.

\section{Conclusions}
In this paper, we proposed a coordinated routing strategy to improve bus schedule adherence while enhancing CAV travel efficiency in mixed-traffic environments. The developed approach proactively identifies a minimal set of CAVs for rerouting based on real-time traffic monitoring conditions and anticipated potential congestion. Simulation experiments conducted in SUMO based on realistic urban road network data demonstrate that the proposed strategy improves both transit reliability and CAV performance across diverse traffic scenarios. A potential direction for future research is to focus on extending this framework to accommodate multi-modal transportation systems and assessing its scalability in large-scale urban road networks.

%%%%%%%%%%%%%%%%%%%%%%%%%%%%%%%%%%%%%%%%%%%%%%%%%%%%%%%%%%%%%%%%%%%%%%%%%%%%%%%
%-----------Section VII. Acknowledgment
%\section{Acknowledgment}
%The authors would like to thank 
%%%%%%%%%%%%%%%%%%%%%%%%%%%%%%%%%%%%
\bibliographystyle{IEEEtran}
\bibliography{ACC_2026,IDS}

\end{document}